# What Happens to Intentional Concepts in Requirements Engineering If Intentional States Cannot Be Known?

Ivan J. Jureta[1]

[1] Fonds de la Recherche Scientifique - FNRS & Namur Digital Institute, Université de Namur
`ivan.jureta@unamur.be`

**Abstract.** I assume in this paper that the proposition "I cannot know your intentional states" is true. I consider its consequences on the use of so-called "intentional concepts" for Requirements Engineering. I argue that if you take this proposition to be true, then intentional concepts (e.g., goal, belief, desire, intention, etc.) start to look less relevant (though not irrelevant), despite being the focus of significant research attention over the past three decades. I identify substantial problems that arise if you use instances of intentional concepts to reflect intentional states. I sketch an approach to address these problems. In it, intentional concepts have a less prominent role, while notions of time, uncertainty, prediction, observability, evidence, and learning are at the forefront.

**Keywords:** Requirements Engineering, Goals, Intentionality, Foundations.

## 1    Introduction

In this paper, I assume that the proposition "I cannot know your intentional states" is true. I call this the *Non-Verifiability Proposition (NVP)*. I analyze its consequences on conceptualizations used for representation and reasoning about system requirements.

What does it mean exactly that I cannot know your intentional states? Here, it means that I cannot know accurately what you want, need, desire, or more broadly, your exact emotions, moods, beliefs, etc. I can talk and make assumptions about your intentional states. Based on communication and other cues, I may believe to have evidence that some of my assumptions are true. But ultimately, I have no reproducible way to ascertain in a clear-cut way if any of those assumptions I have about your intentional states are right or wrong. All I have is my own internal cognitive apparatus. While I may be making these assumptions, and doing the usual things we do when thinking about what others might think, I cannot ascertain that I am right or wrong.

The reason NVP matters is that so-called "intentional concepts" have a central role in Requirements Engineering (RE) research for several decades now. The basic and persistent idea in RE is that requirements need to reflect the purpose of the system-to-be, and that purpose originates in the intentional states of system stakeholders. Roughly, if stakeholders desire that system does something, then we will carry this over to a requirements model as an instance of an intentional concept, namely "goal", and that goal will one of the goals that the system-to-be should be engineered to achieve.

In this paper, I am concerned with what happens to intentional concepts if we assume that intentional states cannot be known. My argument is structured as follows:

1. I start from the observation that intentional concepts, such as "goal", "desire", "intention", "belief", have played, and continue to play a central role in many research contributions in Requirements Engineering for almost three decades so far. **Section 2** identifies some of these concepts, recalls their role and common definitions in RE research, and what they have been proposed for in RE practice.
2. I argue that if you use intentional concepts in RE, then you must make additional assumptions. I call these Requirements Intentionality Assumptions (RIA). I identify them in **Section 3** and relate them to long-standing topics in philosophy on folk psychology and social cognition, and on metarepresentations in psychology and linguistics.
3. In **Section 4**, I identify six problems which arise if you take NVP to be true and you want to use intentional concepts for RE. I argue that these problems exist when instances of intentional concepts are used to convey assumed intentional states. I further argue that the presence of these problems leads to the conclusion that instances of intentional concepts give low quality requirements.
4. Assuming you agreed with me that the problems identified in Section 4 should be taken seriously, or that you at least remain open to further debate, I use **Section 5** to sketch an alternative requirements concept to address the problems.
5. In **Section 6**, I discuss limitations, summarize conclusions, and identify some of many open questions.

Motivation for this paper comes from the tension I experience between the research I do in RE and my practice of it as participant in software product design and development teams over the last ten years. I have helped design and launch about a dozen software products and businesses. Despite my own arguments in RE research in favor of intentional concepts, some of these ventures have suffered from taking requirements to convey predominantly intentional states. It looked consequently worth exploring the consequences of NVP being true.

## 2    Intentional Concepts in Requirements Engineering

This Section uses well-cited prior research to recall common ideas on intentional concepts in RE, and the "goal" concept specifically. The goal concept and the ideas around it have been so influential that there is Goal-Oriented RE, a field on its own within RE. In a survey of Goal-Oriented RE [1], we have the following.

> "A goal is an objective the system under consideration should achieve. Goal formulations thus refer to intended properties to be ensured; they are optative statements as opposed to indicative ones, and bounded by the subject matter.
>
> Goals may be formulated at different levels of abstraction, ranging from high-level, strategic concerns (such as 'serve more passengers' for a train transportation system or 'provide ubiquitous cash service' for an ATM network system) to low-level technical concerns (such as 'acceleration command delivered on time' for a train transportation system or 'card kept after 3 wrong password entries' for an ATM system).

[...] The system which a goal refers to may be the current one or the system-to-be; both are involved in the RE process. High-level goals often refer to both systems. The system-to-be is in essence composite; it comprises both the software and its environment and is made of active components such as humans, devices, and software. As opposed to passive ones, active components have choice of behavior; henceforth we will call them agents. Unlike requirements, a goal may in general require the cooperation of a hybrid combination of multiple agents to achieve it.

[...] Goal identification is not necessarily an easy task. Sometimes they are explicitly stated by stakeholders or in preliminary material available to requirements engineers. [...] In our experience, goals can also be identified systematically by searching for intentional keywords in the preliminary documents provided, interview transcripts, etc."

In everyday language, a goal is the object of a person's ambition, something that they act to achieve. In Goal-Oriented RE, there are goals of (or held by) people or systems. A system can have a goal in the sense that the system, with all the machines, software and people involved, should achieve that goal. Goals come from people related to the system. They are generically called "stakeholders" in this paper.

But are these goals used to reflect intentional states of system stakeholders? If yes, then these goals are what is called "intentional concepts"; if not, they are something else. Consider that question more carefully and four cases turn up.

In **Case A**, I say that the system should do something because I want it, and I want the system to do it with me, or on its own for me. I want the automated thermostat to regulate temperature in my home based on my past settings, inside and outside temperature, air quality, etc. In this case, it is me who wants that, and the system will have that goal because I had that desire, and I want the system to be designed to satisfy that desire. Case A is me who defines system's goals to reflect, and because of, my own understanding of my own intentional states. In this case, I am the requirements engineer and I am the system stakeholder. This makes for a short route - relative to other cases below - from intentional states to requirements. Clearly, in this case "goal" is an intentional concept.

In **Case B**, I may prefer to deal with my thermostat without the system, but I happen to be the one who should design requirements for a system which regulates home temperature automatically on behalf of its users. In this case, I conclude through requirements elicitation that there are people who want to have temperature automatically regulated. It is not me who is in or has the intentional states which are reflected in the goal. It is someone else. I believe they have these intentional states, and since I should design a system to act accordingly, I give that system the goal which reflects these intentional states. Case B is a longer route (than Case A) from intentional states to requirements. I form beliefs about stakeholders' desires, and I give the system goals to reflect desires.

In Case A and Case B, goals are intentional concepts, owing their content and existence to the need to capture something about intentional states.

Now, you may argue that goals are nothing but a different and more practical name for (some types of) requirements, and that there is no need to worry so much about their relationship to intentional states. That is a possible reading of Goal-Oriented RE. We

4could say that it is more useful to get requirements from talking about goals, and that this is because system stakeholders may find goals to be a more convenient conceptualization, a better abstract tool, or easier to talk about, when trying to identify what a system should do with and for them.

Let us call that reading *Light Intentionality*. It consists of using nouns such as "goal" to talk about requirements, while at the same time not explicitly relating them to intentional states. There are problems with that view. Take some goal instance and ask why it exists in a requirements model or specification. The likely answer is that someone wants the system to do something and the goal represents that. If the goal reflects stakeholders' goals, then you are talking about intentional states.

There is a way to distance oneself from intentional states and hold on to Light Intentionality. It consists of focusing not on intentional states as the origin of goals, but on what stakeholders communicate. This Case C is the view from Zave and Jackson [2], who argued that requirements come from "optative statements":

> "Statements in the 'optative' mood describe the environment as we would like it to be and as we hope it will be when the machine is connected to the environment. Optative statements are commonly called 'requirements.'"

In this third case, we find goals or requirements in optative statements.[1] That is, she is the recipient of communication, and treats optative statements as a cue to document goals which reflect what the optative statements are conveying.

So far, in Case A and Case B, an instance of a goal concept is a representation of something about intentional states. In Case C, a goal instance is a representation of something communicated in a specific way.

Observe that you could hold the view that there are no substantial differences between Case B and Case C, if you assume that communication reflects intentional states. This is an argument Mylopoulos, Faulkner, and I presented in the so-called Core Ontology for Requirements [3]: there, we said that there are intentional states, they are conveyed by system stakeholders to requirements engineers via communication, that the requirements engineer look for speech acts to identify instances of different types of information used when doing RE. For goals specifically, we said that they are a record of desires expressed through specific kinds of speech acts.

The strongest relationship of intentional states to requirements seems to be made in Yu's iStar modeling language [4]. This is the foundation for Tropos [5], and was an important influence on Techne [6]. In iStar models, goals always belong to agents. Some agents are people, others (parts of) machines (software and hardware). The language has more than goals, with its notion of intentional dependency, tasks, and the modeling of the rationale of agent's actions using intentional concepts: i.e., agents do things because they hold goals, know means to achieve goals, and these means are tasks they can perform and resources they have access to. An important notion in iStar is that we are modeling the rationale that agents have for collaborating with others, using intentional concepts.

---

[1] As a side note, Zave and Jackson's requirements are van Lamsweerde's goals (even if van Lamsweerde's uses "requirement" as something else, namely a goal which a single agent is responsible for; that makes no difference in this paper).

If Case C is Light Intentionality, then Case A and Case B fall under something we can call *Hard Intentionality*, where requirements are closely tied to intentional states.

If unhappy with these cases, you could argue that whatever intentional states may be, we should focus on understanding how the system-to-be will be used and we can leave the debate of the why as somehow separate. Which business processes, use cases, scenarios, standard operating procedures should the system do or support? This is Case D, in which - for the purposes of this paper - I put all scenario and process-based conceptualizations of requirements. There, requirements reflect ways of doing things.

But isn't Case D Light Intentionality too? If the system should be made to do as is done now, those current processes are still done for some purpose, and that purpose must have been desired, wanted, needed, etc. by someone at some point in the past. If the system needs to do something in ways not done now, these new ways reflect, again, what someone must be wanting, needing, etc. Case D looks like Light Intentionality.

The role of the notion of system purpose is in fact critical in RE. A major change, as far as I can tell, between system design and engineering before RE set itself up as a discipline of research and industry, and after, is that with RE, we should first identify and specify the system's purpose before anything gets made.

The discussion in this section circles back to the observation that requirements need to ensure a system is made to fit its purpose, and that purpose originates in what is wanted, needed, desired, etc. by those who have a say when that system is being designed. Seems non-controversial.

To be clear, not all requirements originate in a system's purpose, some originate in the properties of the environment where the system should run. These properties can be neutral to intentional states: gravity on Mars is weaker than on Earth, whatever we may desire about that. But these neutral environment properties still should, so to speak, go through intentional states of stakeholders: what if I need to design a shuttle to land on Mars for stakeholders who do not believe that gravity there is different from here? So not only goals matter, there are other intentional states to consider. The point here is that, again, intentional states matter for RE.

## 3  Requirements Intentionality Assumptions

Goal-oriented RE, especially in Yu's work and our own Core Ontology for Requirements and Techne, uses the language of folk psychology. That language appears clearly elsewhere in computer science, in some strands of artificial intelligence. Bratman [7] suggested an explanation of rational behavior through notions of belief, desire, and intention. The so-called BDI model [8] was one of the foundational ideas in research on Multi-Agent Systems for instance. Intentional notions play an important role in research from Levesque [9], Halpern [10], and others on knowledge, belief, awareness, etc., in a computational context. What is folk psychology?

> "Folk psychology is a network of principles which constitutes a sort of common-sense theory about how to explain human behavior. These principles provide a central role to certain propositional attitudes, particularly beliefs and desires. The theory asserts, for example, that if someone desires that p, and this desire is not overridden by other desires, and he believes

> an action of kind K will bring it about that p, and he believes that such an action is within his power, and he does not believe that some other kind of action is within his power and is a preferable way to bring it about that p, then ceteris paribus, the desire and the beliefs will cause him to perform an action of kind K. The theory is largely functional, in that the states it postulates are characterized primarily in terms of their causal relations to each other, to perception and other environmental stimuli, and to behavior." [11], [12]

Using intentional concepts for RE means subscribing to the ideas outlined above. It means seeing human behavior in a certain way, as propositional attitudes and causal links between them. It means, for example, that actions are explained in terms of an interplay of beliefs, desires, choices, and commitments.

For illustration, consider how Cohen and Levesque [9] use the language of folk psychology when defining the notion of "intention" of a computational agent.

> "Intention will be modeled as a composite concept specifying what the agent has chosen and how the agent is committed to that choice. First, consider the desire that the agent has chosen to pursue as put into a new category. Call this chosen desire, loosely, a goal. By construction [in their formal framework], chosen desires are consistent. We will give them possible world semantics, and hence the agent will have chosen a set of worlds in which the goal/desire holds. Next, consider an agent to have a persistent goal if he has a goal (i.e., a chosen set of possible worlds) that will be kept as long as certain conditions hold. [...] Persistence involves an agent's internal commitment to a course of events over time. Although a persistence goal is a composite concept, it models a distinctive state of mind in which agents have both chosen and committed to a state of affairs. We will model intention as a kind of persistent goal."

The first quote in this Section illustrates the language that folk psychology uses for describing, explaining, and making predictions of human behavior. The second quote is an illustration of how that language becomes a tool for motivating, describing, and explaining design choices when we make formal models of computational agents.

Folk psychology is used in the same way in RE. We need concepts and relationships, and we need to construct specialized (formal or not) languages for the representation and reasoning about requirements. We can use the language of folk psychology to motivate and justify our designs of these abstract toolsets. For instance, we can say that goal instances capture conditions which are desired by stakeholders, which amounts to defining goals by appealing to intentional states. We can then say that goals can be inconsistent if there are beliefs that they cannot be achieved together; we can define a concept, such as "domain assumption" in the Core Ontology for Requirements, to capture conditions that are the object of such beliefs. By drawing on folk psychology, we can construct conceptualizations of requirements which are inspired by and grounded in various folk psychology ideas. As discussed in Section 2, this has already been done in RE, specifically in Goal-Oriented RE.

If we do that in RE, we should assume that there is something to folk psychology. That at the very least it is a useful language for describing internal dynamics of individual human behavior as well as of social or collective behavior, even if that language may prove to be deficient in constructing valid explanations of human behavior, as Churchland [13] and others have argued (see debates in, e.g., [11], [12]).

If requirements capture intentional states, then intentional states justify the presence of requirements.

The first part of Requirements Intentionality Assumptions (RIA hereafter, both in singular and plural) are, then, assumptions that folk psychology makes about mechanisms producing human behavior. It is, for example, that actions are a function of beliefs and desires.

The second part of RIA is that the language used in folk psychology is useful for describing, explaining, predicting, generalizing patterns of human behavior and its causes. This means that it is useful to talk of desires, beliefs, intentions, and such, when communicating with others about human behavior.

The third part of RIA is the proposition that we can explain why there is such and such requirement on grounds of there being, and being communicated, intentional states of those who have a stake in the system-to-be. That is, some requirements exist because some intentional states exist and have been conveyed by stakeholders to requirements engineers.

The fourth part of RIA is that we should talk about requirements of systems using folk psychology concepts, that there are important relationships between folk psychology concepts and requirements concepts. They are connected in that it is unclear what the latter are, if we untie them from the former.

What I have argued so far, is that if you use intentional concepts in RE, then you should take either the Weak Intentionality or the Hard Intentionality stance. More importantly, whichever of those two you take, you should take RIA seriously. You can see RIA as that batch of ideas that tie conceptualizations of requirements to the broad and malleable folk psychology conceptualizations of human behavior.

Accepting RIA seems indeed to make sense when doing RE, if we see RE as a collective activity where the ability of a requirements engineer to identify (the right) requirements depends strongly on their ability to "read other people's minds", that is, to make assumptions about others' intentional states. The following passage is a neat description of what seems to be going on when RE is done too, even if it is taken from a contribution to cognitive science on how intentional states may be shared.

> "[H]uman beings, and only human beings, are biologically adapted for participating in collaborative activities involving shared goals and socially coordinated action plans (joint intentions). Interactions of this type require not only an understanding of the goals, intentions, and perceptions of other persons, but also, in addition, a motivation to share these things in interaction with others - and perhaps special forms of dialogic cognitive representation for doing so. The motivations and skills for participating in this kind of 'we' intentionality are woven into the earliest stages of human ontogeny and underlie young children's developing ability to participate in the collectivity that is human cognition." [14]

All that I said so far in this section may seem like common sense or common knowledge. Folk psychology, despite vocal critics (Churchland, Stich and others), remains an important conceptualization of individual human behavior (i.e., an important body of theories of mind), as well as an important part of models of how people collaborate. It is also taken seriously in pragmatics.

> "Pragmatic studies of verbal communication start from the assumption [...] that an essential feature of most human communication, both verbal and non-verbal, is the expression and recognition of intentions. On this approach, pragmatic interpretation is ultimately an exercise in metapsychology, in which the hearer infers the speaker's intended meaning from evidence she has provided for this purpose. An utterance is, of course, a linguistically-coded piece of evidence, so that verbal comprehension involves an element of decoding. However, the decoded linguistic meaning is merely the starting point for an inferential process that results in the attribution of a speaker's meaning." [15]

The case for making the assumptions in RIA seems strong, if we take it that RE involves communication in the above sense, that it involves collaboration, and that requirements reflect intentional states.

Taking intentional concepts in RE seriously means, I have argued, taking RIA seriously. And that means subscribing to many other things, namely specific theories of mind, collective action, and communication. As I illustrated above, intentional concepts are a gateway to theories which seem to be widely shared in their respective research communities. Subscribing to them does not look controversial. You could even argue that it is not clear how we would talk about requirements, if we could not speak in terms of what stakeholders may think, want, need, believe, or know.

## 4       Problems

According to NVP, I cannot know your intentional states. In more precise terms, I have no reproducible way to verify if my assumptions about your intentional states are true. "True" here means correspondence of what I assume your intentional states to be to the actual intentional states you are in. The best I can do is have assumptions about your intentional states, and search for cues in communication if I am right or wrong [15].

If I see you ordering a cup of coffee with a friend, I may believe that you want to drink coffee. But you may in fact be ordering coffee as a courtesy to your friend who wanted to have a chat over coffee. Did you desire coffee, or was it the chat with the friend, or both, or something else? How can I find evidence that supports my assumptions about why you want to have that coffee? I can ask you, I can observe, I might ask others their beliefs about your intentional states, infer from answers and observations, update my current beliefs, and then do it all over again. Looking for evidence this way is consistent with assumptions in RIA.

But if you take NVP to be true, then there is no procedure which can move you from assumptions about others' intentional states to knowledge of their intentional states. No evidence will ever get you to know my true intentional states. While that may seem a strong statement, it is consistent with criticisms of the existence of intentional states for example. It is not a new idea.

Why does that matter for RE? It matters because if you use intentional concepts and you take NVP to be true, then you should conclude that you can never know that you got the right requirements from stakeholders. The right ones being those which accurately reflect their true intentional states. Indeed, you cannot know stakeholder's intentional states. Therefore, instances of your intentional concepts are at best your bets that you got their intentional states right (and that you got metarepresentations [16] of those intentional states right). If they are not the right requirements, then solving these requirements, i.e., designing a system which is expected to satisfy them, will give a system that is solving the wrong problem.

If, instead, you use intentional concepts and take NVP to be false, then you seem to be capable of actual mind-reading. I will not take line of argument seriously in the rest of the paper.

One reason why NVP should be taken seriously, and intentional concepts less than they are, is that this makes us recognize that instances of intentional concepts do not correspond and do not reflect something we know. At best, they reflect what we believe to have understood that stakeholders were communicating. We fall back to the Zave and Jackson's idea that requirements are due to optative statements.

Unfortunately, it gets worse. Remember that above, the "right" requirements are those that fit stakeholders' true intentional states. It turns out that these are probably not the right requirements.

When we ask stakeholders what they want, or if we introspect what we may want from a system (or what they may want), we make assumptions not about our (their) current intentional states, but about our (their) future intentional states. Looks like a lot of mental work, especially when there is innovation in the system-to-be.

If you ask me what I expect from a driverless car (which I have not experienced at the time of writing), I must tell you what I will desire in that future, in which I will have that driverless car that you are designing for me. I must imagine that future. I must imagine myself in it. I must imagine that which you have yet to design and build for me, and I must imagine how I would feel about it, what I would believe about it, and so on. Notice that as I do so, while operating with only bounded rationality [17], I am focusing on imagining an isolated experience (being driven in a driverless car), and am disregarding any other experience which I could have also imagined, and which may be related to my imagined intentional states about that system. What if I also imagined that driverless cars were in fact moving not on roads, but in the air, yet I am afraid of flying?

Now that's a mess. If I cannot know your current intentional states, can I know better your future intentional states? Do you know better your future intentional states? From personal experience, I cannot claim to know what I will like or dislike in a few months or years. Can you?

It is worse than that if there is innovation in the system's design. Not only am I asking you about your future intentional states, I am asking you about your future intentional states about phenomena which you have never experienced. Would you have known in 1995 if you would desire an iPhone made in 2017? Would you have known in 1995 if you would prefer one button or two buttons on the front face of a smartphone? Could you have experienced a smartphone in 1995? No. Unless you just came back from the future. As Marty McFly says in a "Back to the Future" movie: This is heavy.

Taking NVP seriously leads to the conclusion that if we want to get requirements from intentional states, we are trying to get them from current intentional states which are about hypothesized future intentional states, which themselves can be about phenomena we may never have experienced. What are, then, the right requirements?

The unavoidable conclusion is that there are several substantial problems which arise if we are looking for the right requirements in intentional states.

- **Non-Verifiable Intentionality Problem:** Instances of intentional concepts are about intentional states, yet NVP tells us we cannot know these intentional states.
- **Uncertain Intentionality Problem:** Because the system-to-be does not exist yet at the time its requirements are being specified, instances of intentional concepts are about future intentional states. These seem even less amenable to being known than present intentional states. These are not intentional states experienced when requirements are elicited.
- **Speculative Intentionality Problem:** Because the system-to-be may generate new experiences, and these experiences are not known by system stakeholders at the time stakeholders communicate their intentional states, instances of intentional concepts are about intentional states that would arise in the future if stakeholders in that future have lived these presently unknown experiences.

There is more bad news. Time passes from when requirements are elicited and specified, to when the system-to-be is made and runs and can be experienced (used) by its stakeholders. Intentional states conveyed at requirements elicitation time may have expired, or at least changed enough that they are in fact no longer satisfied (to the same extent) by the system-to-be. I call this the **Expiring Intentionality Problem**.

Unfortunately, it is too optimistic to see only the four problems raised so far. What if what is said is not entirely true? Or more broadly, what if there is no reason to trust that what is communicated genuinely reflects true intentional states? What if both the mode and content of communication of intentional states are distorted, for example, to provide partial information? What if what is said is, so to speak, cheap talk (i.e., it is said but has no bearing on what is done)? I call all such cases the **Distorted Intentionality Problem**; there are interesting nuances between these cases, but are not important for this paper.

Even if there are only the best intentions, people forget to mention information which may turn out to be critical for making sense of intentional states that they may be in. They may omit important conditions, observations, remarks. They may fail to mention what is obvious to them, even if it may not be apparent at all to aliens coming from other domains. This is called the **Incomplete Intentionality Problem** in this paper.

In short, requirements which originate in intentional states come together with six problems: non-verifiability, uncertainty, speculation, expiry, distortion, and incompleteness.

Do these problems matter? The success of RE has two dimensions. One is that the system runs according to its requirements specification. The other is that it meets expectations of its stakeholders. These are called, respectively, engineering quality and service quality (as in quality of services that the system provides to its stakeholders).

Engineering quality can be high, yet the delivered system may fail to deliver service quality. This is because engineering quality is fitness to a specification, while service quality critically depends on how well the specification satisfies the right requirements. If the six problems are present, we cannot know if we have the right requirements. The specification, even if precise, comprehensive, and clear, will involve the risk that the system satisfies the wrong, or perhaps only a subset of the right requirements.

As an aside, note that this risk is solved partly through RE methods, and partly by freezing and requiring formal approval of the frozen the specification. But freezing the specification does not imply that intentional states will also be frozen.

## 5   Sketching a Solution

By demoting intentional states among lesser sources of requirements, we should answer two questions. Where do we preferably get requirements from, if not from intentional states? How do these other sources deal with problems identified in Section 4?

The first step is to avoid the language of folk psychology and use the word "requirement" instead of "goal" for top-level, most abstract requirements. The next step is to identify necessary properties for the requirement concept.

*Non-Verifiable Intentionality Problem* exists because the content of a requirement is a statement about something that cannot be observed, and thus, whose relationship to that statement cannot be reproducibly verified. We therefore need requirements to be statements which are verifiably reproducible. This also means that the statement that makes the requirement should be treated as a hypothesis which needs to be verified. It follows that, for each requirement we need to answer the following questions.

- **Hypothesis:** Which hypothesis is/are stated in the requirement?
- **History:** How was the hypothesis formulated? Where did the hypothesis originate?
- **Verification:** How can the hypothesis be reproducibly verified?
- **Instrument:** What is used to establish the hypothesis-experience relationship?

In a venture where I took part, a high-level requirement was that the system will remove the bias that brokers introduce in the price of road freight transportation. The hypothesis is that prices paid by shippers to carriers, via the system, will be systematically lower than historical broker-set prices for comparable shipments (similar freight, routes, etc.). The entire system was in fact designed to verify that hypothesis. The history of the hypothesis was the observation by the founders, that freight brokers charge inconsistent prices for similar freight and routes, that is, that prices do not reflect market conditions, but broker's ability to exploit information asymmetries. Verification amounts to the entire lifecycle of the system, while the system itself is the instrument. Another requirement, which has a more modest scope, was that market participants will perform the onboarding of their resources (freight, equipment, drivers, etc.) by themselves. The history was that it requires less resources to have the customers provide, by themselves, the required information about resources. Verification process involves building a prototype of the onboarding process, then having a small sample of select prospects use that prototype. The prototype was only part of the instrument, the other part being a questionnaire filled out by observing how customers did it, and by interviewing customers about the experience of using the prototype.

For verification to be reproducible, its steps and instruments need to be accessible to others. It follows, for example, that neither one's own mind can be the instrument, nor one's own thinking the verification process. This makes prototypes more interesting instrument than, say, a textual specification.

*Uncertain Intentionality Problem* exists because requirements are about the future. The problem with the future is that we must wait to experience it. Waiting costs. If what we say about the future matters today, then our current actions depend on that which we expect in the future. A requirement thus involves the risk of being a wrong prediction. That risk is proportional to the cost of waiting to verify the prediction. It follows that we should minimize the time to verify the prediction. A requirement's age is thus an input to the estimation of the risk it carries of being wrong. **Age** and **Risk** become necessary properties of any requirement. Risk is in two parts, as it is necessary to describe the outcomes expected in case the hypothesis does not fit experience, and a description (probability?) that this happens. If waiting costs, then risk will be a function of a requirement's age.

*Speculative Intentionality Problem* can be addressed by approximating to stakeholders sooner the unknown future experience that they speculate about. Stakeholders should learn requirements, rather than get them from speculation, through use of instruments which approximate the new experience. Moreover, this needs to happen in a proxy environment which provides conditions resembling as much as possible the conditions in which the stakeholders expect to use the system-to-be. Requirements should originate in observable phenomena which requirements engineers can experience in similar ways in which stakeholders already do. It is practices and conditions in an environment that clash with the elusive internal ambitions and motives of stakeholders, leading them to have requirements in the first place.

When trying to learn requirements for a system which should match shippers in need of freight transportation, with carriers, go to shippers to do the work they do now, and go to carriers to do what they do, and finally, go to brokers who mediate most transactions in the freight transportation market. When trying to design a system which uses AI to provide instructions for improving one's running, take time to become even a novice running coach and work under an expert. This is hard to do, it takes significant time, it requires interested parties in proxy environments, who are willing to open their practices to newcomers. It also places a greater burden on requirements engineers, who can no longer take interviews and distant observation too seriously, but have to apply themselves to doing that which in fact generates requirements in the first place.

To reduce speculation, requirements engineers learn by doing stakeholders' actions in the proxy environment, while stakeholders learn through changing their past actions using instruments designed to validate hypotheses. Each requirement comes with its own set of **Practices** that requirements engineers need to be interested in, in that the system will rely on and change these practices, and **Proxy Environments** in which these practices can be realized in approximate conditions.

*Expiring Intentionality Problem* is due to the unknown lifetime of a requirement. This is addressed in several ways. One is by minimizing the Age of the requirement by which we do a first verification of the hypothesis. Another is by repeating verification, if we observe cues that the requirement may be expiring. In the earlier example, of the hypothesis that market participants will on-board by themselves their resources, an **Expiry Cue** was that some of the stakeholders asked for features that would enable the operations department of the firm to on-board customers' resources on their behalf (suggesting that stakeholders are predicting that not all customers are willing to perform on-boarding on their own).

To address the *Distorted Intentionality Problem*, there are distortion cues to monitor for. For example, inconsistencies in the information that stakeholders provide are such a cue. But the main direct means to identify distortion is, as above, early verification, since it confronts stakeholders with an approximation of their future experience. Same applies to the *Incomplete Intentionality Problem*, in that there are cues to look out for, while early verification should help stakeholders identify incompleteness.

## 6  Discussion and Conclusions

At best, this paper is an exploratory identification and analysis of six problems which arise if intentional concepts are used to represent stakeholders' intentional states, under the assumption that intentional states are an important source of requirements. The arguments laid out are theoretical claims, supported only by personal experience, but not by rigorous reproducible research. Despite these important drawbacks, the argument is worth considering, given that it reaches novel conclusions, yet starts from non-controversial and well known premises.

I started from the assumption that I cannot know your intentional states. This led me to identify a set of properties of a "requirement" concept which distances itself from intentional concepts. To the best of my knowledge, the properties have not been tied to the requirement concept in the past, and have not been used to describe intentional concepts in RE.

The paper raises many new and interesting questions. What does it mean exactly to learn requirements? How can that be facilitated? How do we elicit information useful for the formulation of the hypotheses? How can we verify these hypotheses, and what are the merits and limitations of alternative verifications? What are the consequences of the need for early verification on how we refine requirements? Which instruments to use for verification, and how do they compare? And so on.